\documentclass[preprint,showpacs,preprintnumbers,amsmath,amssymb]{revtex4-1}

\usepackage{graphicx}
\usepackage{color}
\usepackage{longtable}

\begin{document}

\title{Crystal Structure and Magnetic Properties of the New
Zn$_{1.5}$Co$_{1.5}$B$_7$O$_{13}$Br Boracite}
\author{Roberto Escudero}
\email{escu@servidor.unam.mx}
\author{Francisco Morales}
\affiliation{Instituto de Investigaciones en Materiales, Universidad
Nacional Aut\'{o}noma de M\'{e}xico. A. Postal 70-360. M\'{e}xico,
D.F., 04510 M\'EXICO.}
\author{Marco A Leyva Ramirez}
\affiliation{Departamento de Qu\'{\i}mica, Centro de Investigaci\'{o}n y
Estudios Avanzados del IPN. 07360, M\'{e}xico, D. F.}
\author{Jorge Campa-Molina}
\author{S. Ulloa-Godinez}
\affiliation{C. Universitario de Ciencias Exactas e Ingenierías,
Universidad de Guadalajara, Laboratorio de Materiales Avanzados
Departamento de Electronica. 044840, Guadalajara, Jalisco. M\'EXICO.}

\date{\today}

\begin{abstract}
  New Zn$_{1.5}$Co$_{1.5}$B$_7$O$_{13}$Br boracite crystals were grown
by chemical transport reactions in quartz ampoules, at a temperature
of 1173 K. The crystal structure was characterized by X-ray
diffraction. The crystals present an  orthorhombic structure with
space group Pca2$_1$, (No. 29). The determined cell parameters were:
$a$ = 8.5705(3)\AA, $b$ = 8.5629(3) \AA, and $c$ = 12.1198(4)\AA, and
cell volume,  V = 889.45(5) \AA$^3$ with Z = 4. Magnetic properties
in single crystals of the  new boracite, were determined. The
Susceptibility-Temperature ($\chi-T$) behavior at different magnetic
intensities was studied. The inverse of the magnetic susceptibility
$\chi^{-1}(T)$ shows a Curie-Weiss characteristic with spin $s = 3/2$
and a small orbital contribution, $l$. At low temperatures, below 10
K, $\chi(T)$ shows irreversibility that is strongly dependent on the
applied magnetic field. This boracite is ferrimagnetic up to a
maximum temperature of about 16 K, as shows the coercive field. The
reduction of the  irreversibility by the influence of the magnetic
field, may be related to a metamagnetic phase transition.
\end{abstract}

\pacs{Boracites; Ferromagnetism; Metamagnetic transitions; Schottky
anomaly; Crystal structure}

\maketitle

\section{Introduction}

The term boracites is at present given to more than 25 isomorphous
compounds all with the general formula Me$_3$B$_7$O$_{13}$X, where Me
is one of the divalent metals Mg, Cr, Mn, Fe, Co, Ni, Cu, Zn or Cd
and X is usually Cl, Br or I. Occasionally X can be OH, S, Se or Te
and monovalent lithium substitutes for Me; then the formula becomes
Li$_4$B$_7$O$_{12}$X where X is a halide \cite{levas,nel}. Natural
and synthetic boracites have attracted the attention of researchers
since the early times R. J. Ha\"{u}y \cite{dana,han}. The mineral
boracite Mg$_3$B$_7$O$_{13}$Cl, provides the name of this large
family \cite{nel,dana}. Only other four natural boracites are known:
ericaite, chambersite Mn$_3$B$_7$O$_{13}$Cl \cite{hone}, congolite
Fe$_3$B$_7$O$_{13}$Cl \cite{dana2}, and trembathite
(Mg,Fe)$_3$B$_7$O${13}$Cl \cite{dana2}. Ericaite and trembathite are
natural mixed boracites, which have been taken as motivation on this
investigation to synthesize other types of mixed boracites. The
reason to synthesizing the mixed (Zn,Co)$_3$B$_7$O$_{13}$Br
boracites, is related to the idea to understand more about the
physical and chemical properties, when combining metals. Boracites
have received special attention because of their unusual physical
properties \cite{smart,mathe,caste,campa}. In this contribution we
present results on the crystallographic characteristics and mainly on
the new magnetic properties presented in
Zn$_{1.5}$Co$_{1.5}$B$_7$O$_{13}$Br. We report the magnetic
characteristics from room to low temperatures, investigating the
influence of the magnetic field, and the behavior of the specific
heat at low temperatures. We found that irreversibility in $\chi(T)$
is strongly dependent on the intensity of the applied magnetic field.
We explain this behavior as the transition from low to high spin
changes (metamagnetic transition). So, at low field the material
behaves as a ferrimagnet, and transits to a ferromagnetic state (high
spin) with increasing magnetic intensity. In addition specific heat
measurements show a Schottky anomaly at low temperature. We
interpreted this anomaly as a resonance between the splitting of
spins, from ms states:  3/2 to -3/2, and the thermal energy.

\section{Experimental Details}

\subsection{Crystal growth}

The Schmid  method for the  sample preparation of crystalline
materials was followed \cite{schmid}. The compound was grown by
chemical reaction of vapour phases. Reactants were placed in three
fused quartz crucibles of different dimensions separated by small
quartz rods, and vertically aligned. To facilitate the chemical
transport, the initial compounds (halogenures, metal oxides, and
boric acid) were placed in the bottom, first crucible. The content of
this  was 1.7 g of B$_2$O$_3$ (obtained by dehydrating H$_3$BO$_3$);
the second crucible contains 0.5 g of ZnO and CoO. Finally,  the
upper third crucible contains 0.8 g of each divalent metal halides;
CoBr$_2$, and ZnBr$_2$.   These were placed inside a quartz ampoule
under vacuum. The chemical transport reactions were carried out by
heating the ampoule in a vertical oven  with  the following heating
steps: 1173 K over a period of 50 hours, and 913 K over a period of
20 hours. At the end of this process, once at room temperature  we
observed small single crystals boracites. The single crystals of the
mixed boracites show an intense purple color and approximately  3 mm
in size, those crystals mostly are formed at the end of the lower
crucible.

\subsection{Crystal structure analysis}

X-ray data were collected using graphite monochromated  MoK$\alpha$
radiation $\lambda=0.71073$ \AA\ in a Enraf-Nonius Kappa-CCD
diffractometer. Data reduction was made by the Denzo-Scalepack
software \cite{otwin}. The Structure was solved by direct method and
refined by full-matrix least squares, using the SHELX97 package
software \cite{sheld}.

\subsection{Refinement details}

Refinement of Boracites by single-crystal techniques is complicated
because  the very frequent twins created in the growing process
\cite{abrahams}. In order to solve this complication  we used a twin
matrix method to refine the crystal structure. We detected two
domains  in the studied single crystal by using  the  package PLATON
\cite{spek}. In order to have a better interpretation of the studied
electronic densities a  Multi-Scan absorption correction was used
with the SADABS software \cite{sheld}. Due to high electron density
of the Bromine atoms, the localization of light elements becomes
complicate, and also the corresponding refinement: This fact explains
the low isotropic temperature factor(Uiso) in the Boron and Oxygen
atoms. Accordingly, the X-ray pattern of the studied sample was
refined isotropically. In this case was necessary to consider a
restrain for the occupation. This was taken as 0.5 of Zn/Co atoms;
with this restrain we obtained good agreement for the anisotropical
values for Zn and Co atoms. thus, good convergence and low R value
were obtained.

\begin{table}
\caption{\label{t1}Crystal data and data parameters of
Zn$_{1.5}$Co$_{1.5}$B$_7$O$_{13}$Br boracite.}
\begin{tabular}{|c|c|}
\hline
 Compound & Boracite \\
\hline Chemical  formula & Co$_{1.5}$Zn$_{1.5}$B$_{7}$O$_{13}$Br \\
\hline Formula weight & 550.03 \\
\hline Cryst size [mm] & 0.075 x 0.11 x 0.19 \\
\hline Cryst. system  & Orthorhombic \\
\hline Space group & Pca2$_1$ \\
\hline $a$, [\AA] & 8.5705(3) \\
\hline $b$, [\AA] & 8.5629(3) \\
\hline $c$, [\AA] & 12.1198(4) \\
\hline V, [\AA$^3$] & 889.45(5) \\
\hline Z & 4 \\
\hline $\rho$(calc.), [Mg/m$^3$] & 4.107 \\
\hline $\mu$ [mm$^{-1}$] & 11.366 \\
\hline F(000) & 1038 \\
\hline Index range & $-9 \leq h \leq 10$    \\
& $-11\leq k\leq 11$      \\
& $-15\leq l\leq 13 $\\
\hline $2\theta$ [$^\circ$] & 54.80 \\
\hline Temp, [K] & 293(2) \\
\hline Refl. collected & 10425 \\
\hline Refl. unique & 2027 \\
\hline Refl. observed (4$\sigma$) & 1273 \\
\hline R (int.) & 0.0601 \\
\hline No. variables & 116 \\
\hline W. scheme x/y & $w^{-1} = \sigma ^2F_0^2 + (xP)^2 + yP$ \\
&  $P =(F_0^2 + 2F_C^2)/3 $\\
\hline GOF & 1.001 \\
\hline Final R (4$\sigma$) & 0.0334 \\
\hline Final wR2 & 0.0708 \\
\hline Larg. res. peak [e/\AA$^3$] & 3.140 \\
\hline
\end{tabular}
\end{table}

\subsection{Magnetic measurements}

 Magnetic measurements were performed using a MPMS magnetometer
(Quantum Design). $M-T$ measurements were obtained at three different
magnetic intensities; 100, 1000, and 5000 Oe. The data acquisition
modes were the standard Zero Field Cooling (ZFC) and Field Cooling
(FC). After these measurements we obtained the magnetic
susceptibility as a function of the temperature $\chi(T)$ in terms of
cm$^3$/mol, and the Pascal constant were added. In addition, we
performed isothermal magnetization measurements cycles at various
temperatures.

\subsection{Specific heat measurements}
Specific heat measurements $C_P$ as a function of temperature were
performed using a thermal relaxation method, utilizing a PPMS
(Quantum Design) apparatus. Measurements were performed at low
temperatures with applied magnetic field of 0, 100 and 5000 Oe. The
$C_P$  values were corrected subtracting the addenda due to the
sample support and the grease used to glue the sample on the support.

\section{Results and Discussion}

\subsection{Structural characterization}

\begin{figure}[h]
\includegraphics[scale=0.4]{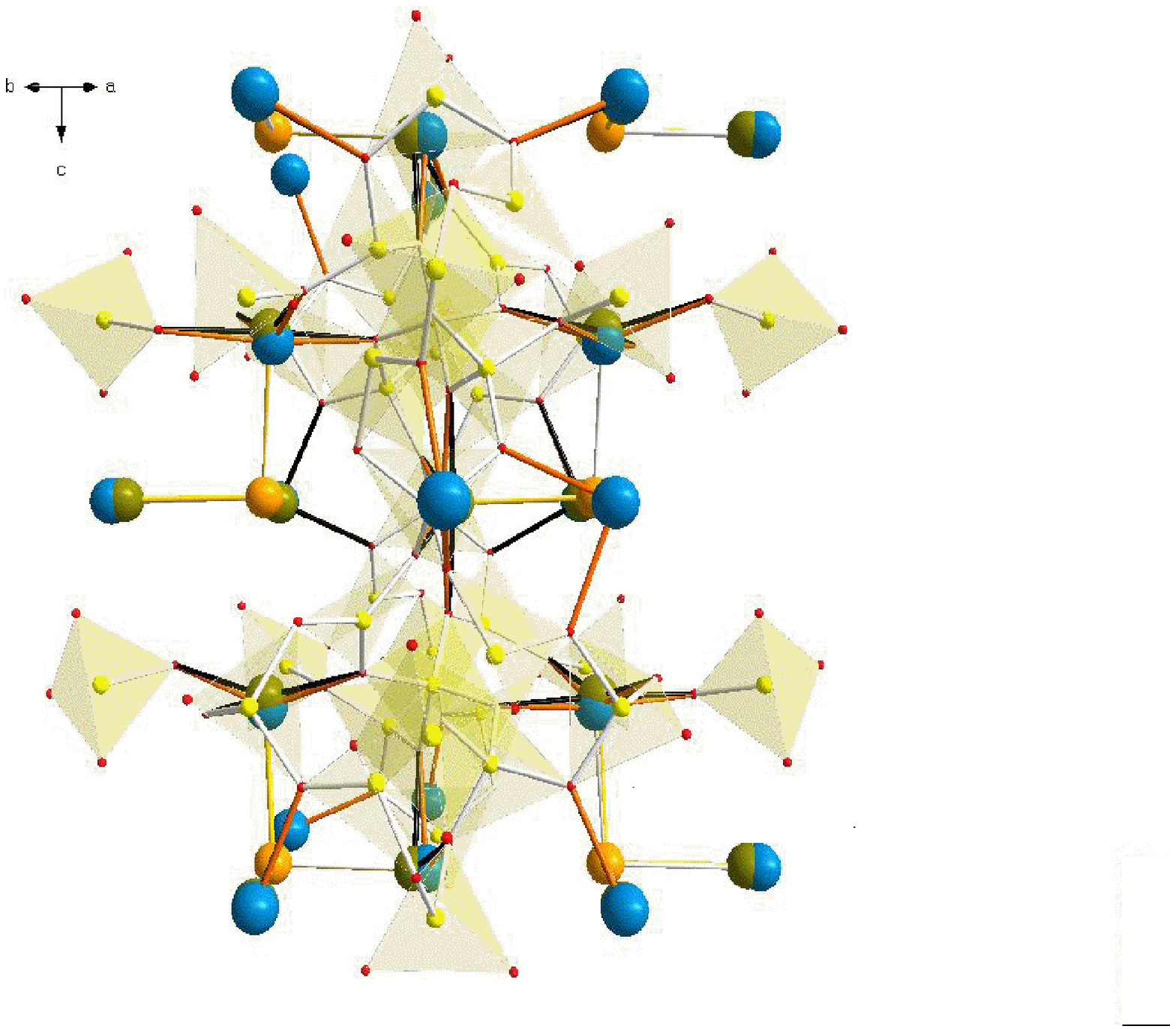}
\caption{(Color online) This figure displays the Unit cell for
Zn$_{1.5}$Co$_{1.5}$B$_7$O$_{13}$Br boracite. Atoms are marked in different
colors and descending in size, as: blue balls, Zn; green balls, Co; orange,
Br; yellow, B; and small red ones, Oxygens.}
\label{fig1}
\end{figure}

X-ray diffraction analysis reveals that
Zn$_{1.5}$Co$_{1.5}$B$_7$O$_{13}$Br compound crystallizes in an
orthorhombic structure with space group Pca2$_1$ (No. 29). The
structural parameters determined are summarized in \ref{t1}. In
Supplementary Material the final refined positional and thermal
parameters are given in  Table 2, and the main interatomic distances
and angles are given in Table 3. The unit cell for this boracite is
shown in \ref{fig1}.

The bond distances and angle values on first approach, looks
abnormal. The reason is because the difference between the
coordination polyhedral bond distance is too big. For instance, the
octahedral polyhedral whose central atoms are Br, is coordinated with
six Zn or Co atoms; the average bond distances Zn(1)-Br and Co(1)-Br
are 2.576(4) and 2.756(5) \AA, which implies a big distortion of the
polyhedrons. This polyhedral distortion is a characteristic of every
boracite compound \cite{ascher,ascher2,thompson,kubel,ito}. We found
that the diagonal value between the lattice constant $a$ and $b$ is
12.1151 \AA, which is almost the same value of the $c$ parameter
(12.1198 \AA). Likewise, observing the bond distance between
different tetrahedra, as BO$_4$, we clearly can distinguish  very
strong deformations. These deformations may provokes that the total
atom charge in the crystal structure  not be neutralized, and
consequently zones with negative and positive charge could appear.
This distortion generates electric dipoles along the preferred
direction of the crystal structure.

\begin{figure}[btp]
\includegraphics[scale=0.7]{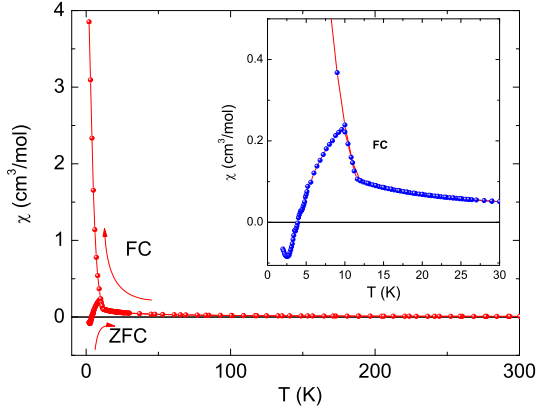}
\caption{(Color online) Magnetic susceptibility vs. Temperature,
$\chi(T)$ of (Zn,Co)$_3$B$_7$O$_{13}$Br boracite single crystal
measured with 100 Oe, in ZFC and FC modes, the arrows show the
increasing/decreasing of the temperature.
At low temperature the  irreversibility is clearly observed. The
insert displays the rapid increase of the susceptibility in the FC
mode, occurring at about 12 K.}
\label{fig2}
\end{figure}

\begin{figure}[btp]
\includegraphics[scale=0.7]{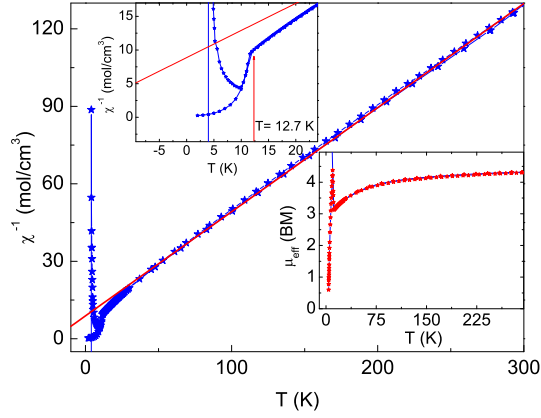}
\caption{(Color online) Inverse of the magnetic susceptibility
$\chi^{-1}(T)$ measured in ZFC and FC modes, at 100 Oe. The continuous
line in the main panel and the top insert shows the fit to the Curie-Weiss
law. Curie constant value was
determined to be  $C = 2.46\pm 0.03$ $cm^3$ K/mol, and $\theta_{CW}$
changes from 19 - 24 K.
The top insert shows an amplification
of the main panel at low temperature. There, the vertical arrow marks the
change of the slope, which is about 12.1 K.  The inferior insert
displays the  effective number of Bohr magnetons, determined as
4.25 $\mu_B$ at room temperature.}
\label{fig3}
\end{figure}

\begin{figure}[btp]
\includegraphics[scale=0.7]{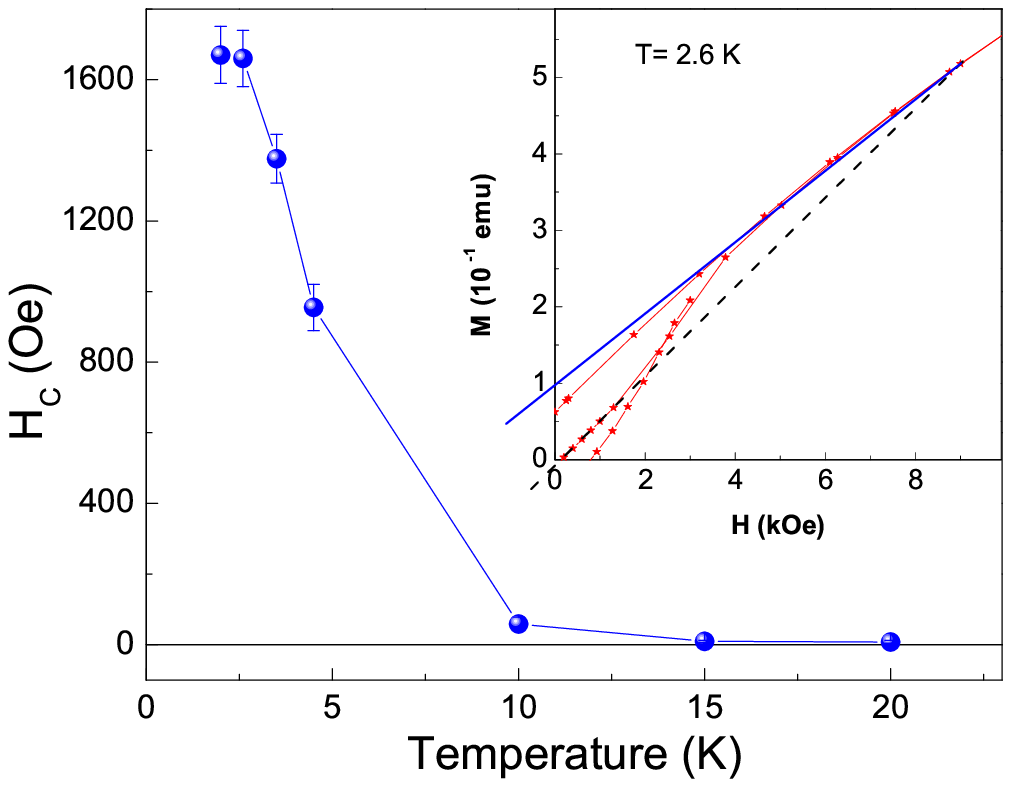}
\caption{(Color online)  Coercive field determined by isothermal
$M-H$ measurements at different temperatures. The ferromagnetic ordering persists
only around  15 - 20 K. The insert shows the influence of magnetic field in the
magnetic properties, This small spin crossover indicates a metamagnetic
transition changing the magnetic ordering from ferrimagnetic to
ferromagnetic.}
\label{fig4}
\end{figure}

\subsection{Magnetic behavior}

\ref{fig2} displays the susceptibility as a function of temperature
measured at 100 Oe. At first glance we can see two characteristics:
In  FC mode $\chi(T)$ presents a paramagnetic behavior at high
temperature, the values are quite small and close to zero. At low
temperature, about 50 K the $\chi$ value gradually increases and  at
about 12 K $\chi$ presents a rapid increase; at 2 K has a huge value
of about 4 $cm^3/mol$. However, in the ZFC mode at 2 K also, $\chi$
is negative but rapidly grows as the temperature rises, the  a
maximum values at 11 K is about 0.23 $cm^3/mol$. This is small in
comparison to the maximum value in FC mode. A that maximum the two
modes present an irreversible behavior when the temperature
decreases. This behavior clearly is displayed in the insert of
{\ref{fig2}. \ref{fig3} shows the inverse susceptibility
$\chi^{-1}(T)$,  used to determine the magnetic characteristics. As
already mentioned, at high temperature the compound follows a
Curie-Weiss law that was fitted with a constant, $C=2.46 \pm 0.03$
$cm^3K/mol$, and an extrapolated Curie-Weiss temperature
$\theta_{CW}$ from $-19$ to $-24$ K. This C value is well fitted with
a total number of Co spins contribution $s$ = $3/2$, and a small
orbital angular contribution $l<1$. It is important to mention that
in orthorhombic crystals (also in cubic crystals) structures the
orbital moment vanishes at first order. In this case the small
deformation of the orthorhombic structure (the angles are not exactly
90 degrees) gives the possibility, as we observed, that the orbital
angular moment $l$ is not totally quenched. In the top insert of this
figure we plotted an amplification of the inverse susceptibility from
2 K  to 20 K. At about 12.7 K an abrupt change of slope occurs. In
the inferior insert we display the effective number of Bohr
magnetons, $\mu_B$, at room temperature are equal to 4.3 $\mu_B$. In
the same figure and at low temperature a ferrimagnetic transition can
be deduced; this ferrimagnetic ordering is according to the negative
$\theta_{CW}$ values obtained by the fitting. nevertheless, as a
confirmation of the ferrimagnetic behavior, and the crossover to high
spin by influence of the magnetic intensity, we studied the
isothermal magnetic $M-H$ characteristics at different temperatures,
from 2 K to 20 K. The coercive field already is observed, and it is
shown in the main panel of \ref{fig4}. It is clear that the coercive
field tends to disappear at temperatures of the order of 15 K. The
insert in this figure also presents one isothermal $M-H$ at 2.6 K to
illustrate crossover from low spin to high spin.

\begin{figure}[btp]
\includegraphics[scale=0.6]{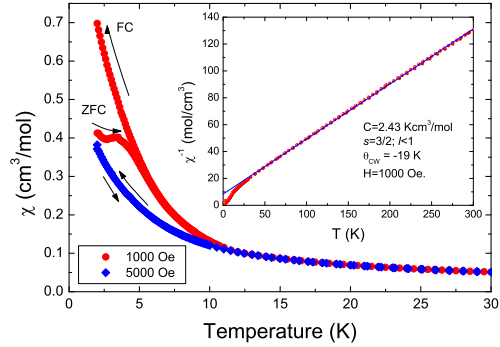}
\caption{(Color online) Main panel presents the magnetic susceptibility,
$\chi(T)$, measured
with 1000 Oe, from 30 K to 2 K.  The characteristic shows important
differences of behavior that are strongly dependent on the intensity
of the magnetic field applied. At higher field, 5000 Oe,  the behavior
is completely reversible, as noted by completed reversibility. This
behavior marks the  metamagnetic transition,
by influence of the applied magnetic field.}
\label{fig5}
\end{figure}

\ref{fig5} displays the susceptibility as a function of temperature
measurements into magnetic fields of 1000 Oe and 5000 Oe. Note that
the irreversible temperature changes at lower temperature and
disappears at 5000 Oe. We interpreted this behavior as  a
metamagnetic transition. However in order to have a clear
confirmation of this feature we plotted at low temperature isothermal
$M-H$ measurements. This characteristic shows a small but discernible
change from low to high spin as the magnetic intensity is increased.
Metamagnetism is due to the applied field strength that overcome the
crystal anisotropic force, producing an abrupt change in the internal
order \cite{hurd}.

\begin{figure}[btp]
\includegraphics[scale=0.3]{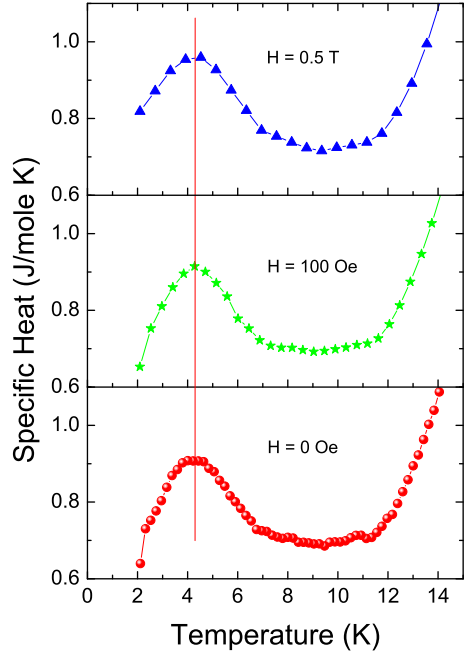}
\caption{(Color online)
Specific heat at low temperature of the boracite
Zn$_{1.5}$Co$_{1.5}$B$_7$O$_{13}$Br  measured at  magnetic field
intensities of 0, 100, and 5000 Oe. The feature displayed at about
4.5 K, clearly can be catalogued as a Schottky anomaly.}
\label{fig6}
\end{figure}

\subsection{Specific heat}

Lastly, in order to confirm the complicated physical characteristics
of this boracite, we studied the specific heat  at  low temperatures.
Our experiments show an anomalous wide peak, that we characterized as
a Schottky anomaly (\ref{fig6}). This occurs at low temperatures,
where spin population may be excited by thermal energy  between $e_g$
and $t_{2g}$ levels of the system. This Schottky anomaly occurs at
about 4.3 K, the thermal energy at this temperature is about $370
\times 10^{-6}$ eV, corresponding approximately to the splitting of
the $\pm s = 3/2$ states.

\section{Conclusions}
A new Zn$_{1.5}$Co$_{1.5}$B$_7$O$_{13}$Br boracite was synthesized by
the vapor transport method. The compound has a orthorhombic crystal
structure, with small distortions, and with the space group Pca2$_1$
(No. 29) and  cell parameters $a$ = 8.5705(3), $b$ = 8.5629(3), and
$c$ = 12.1198(4) \AA, and Volume = 889.45(3) \AA$^3$. Magnetic
measurements performed as a function of temperature and at different
magnetic intensities show that the general behavior at low
temperature changes; first at all,  the irreversibility disappears at
an intensity of about 5000 Oe, and the anomalies change. This
behavior is because the low to high spins crossover, affecting the
general characteristics. We identify this reduction of
irreversibility as a metamagnetic transition.  Lastly, specific heat
measurements at low temperature show an anomaly that clearly was
identified as a Schottky anomaly.

\begin{acknowledgements}

RE thanks CONACyT project 129293, and DGAPA UNAM project IN100711. FM
thanks the partial support of DGAPA UNAM project IN111511. We also
thank to R. Reyes and to F. Silvar for help in technical problems.
\end{acknowledgements}

\end{document}